\documentstyle[12pt,epsfig,axodraw]{article}
\begin{document}
\baselineskip 0.6cm

\def\simgt{\mathrel{\lower2.5pt\vbox{\lineskip=0pt\baselineskip=0pt
           \hbox{$>$}\hbox{$\sim$}}}}
\def\simlt{\mathrel{\lower2.5pt\vbox{\lineskip=0pt\baselineskip=0pt
           \hbox{$<$}\hbox{$\sim$}}}}

\begin{titlepage}

\begin{flushright}
UCB-PTH-04/28 \\
LBNL-56523 \\
\end{flushright}

\vskip 2.0cm

\begin{center}
{\Large \bf Supersymmetric Unification in Warped Space}

\vskip 0.8cm

{\large
Yasunori Nomura
}

\vskip 0.4cm

{\it Department of Physics, University of California,
           Berkeley, CA 94720}\\
{\it Theoretical Physics Group, Lawrence Berkeley National Laboratory,
           Berkeley, CA 94720}

\vskip 1.2cm

\abstract{
Supersymmetric unification in warped space provides new possibilities 
for model building.  I argue that the picture of warped supersymmetric 
unification arises naturally through the AdS/CFT correspondence from the 
assumption that supersymmetry is dynamically broken at a scale $\approx 
(10\!\sim\!100)~{\rm TeV}$, and present several fully realistic theories 
in this framework.  In the minimal model, the bulk $SU(5)$ gauge group is 
broken at the Planck brane by boundary conditions, while supersymmetry is 
broken at the TeV brane.  The theory preserves the successes of conventional 
supersymmetric unification, and yet physics at accessible energies has a 
drastic departure from that of the conventional scenario.  There are also 
several variations of the minimal model giving distinct phenomenologies. 
These theories can provide a basis for phenomenological studies of 
dynamical supersymmetry breaking at low energies. 
}

\end{center}
\end{titlepage}

\section{Introduction}
\label{sec:intro}

Supersymmetry has long been the leading candidate for physics beyond the 
standard model.  It stabilizes the Higgs potential against potentially 
huge radiative corrections, giving a consistent theory of electroweak 
symmetry breaking.  The minimal construction of the supersymmetric 
standard model, which contains supersymmetric multiplets for the 
standard-model gauge and matter fields as well as two Higgs doublets, 
also provides an elegant picture of gauge coupling unification at a scale 
of $M_X \simeq 10^{16}~{\rm GeV}$.  A generic prediction of supersymmetric 
theories -- the presence of a light Higgs boson -- also seems to be 
supported by precision electroweak data. 

Despite all these successes, the construction of a supersymmetric extension 
of the standard model is not yet complete.  The origin of supersymmetry 
breaking is still a mystery and the communication of supersymmetry 
breaking to the supersymmetric standard model (SSM) sector is not yet 
fully understood.  Some good news, though, is that in supersymmetric 
theories, the non-renormalization theorem guarantees that supersymmetry 
can be broken only by non-perturbative effects if it is not broken at 
tree level.  Supersymmetry, then, can be broken at a dynamical scale 
$\Lambda$ of some gauge interaction $G$ responsible for dynamical 
supersymmetry breaking:
\begin{equation}
  \Lambda \sim M_{\rm Pl}\, e^{-\frac{8\pi^2}{|\tilde{b}|\tilde{g}^2}},
\label{eq:dim-trans}
\end{equation}
where $\tilde{g}$ is the gauge coupling of $G$ renormalized at the Planck 
scale, and $\tilde{b}$ ($<0$) the beta-function coefficient for $\tilde{g}$. 
This provides a natural understanding of the smallness of the weak scale, 
as $\Lambda$ is exponentially smaller than the Planck scale for $|\tilde{b}| 
\tilde{g}^2 \ll 8\pi^2$~\cite{Witten:1981nf}. 

What is the scale $\Lambda$?  The answer depends on the mechanism by 
which supersymmetry breaking is mediated to the SSM sector, the sector 
that contains our quarks and leptons and their superpartners.  If the 
mediation occurs through gravitational interactions, $\Lambda \simeq 
10^{10}-10^{13}~{\rm GeV}$, while if it occurs through standard-model 
gauge interactions, $\Lambda$ can be much lower.  These two interactions 
are selected as natural ways of mediation both because they are already 
known to exist and because they give ``flavor universal'' squark 
and slepton masses, which are needed to evade strong experimental 
constraints on the amount of flavor violation beyond that in the standard 
model~\cite{Dimopoulos:1981zb}.  Although it is possible to consider 
a scenario based on mediation through gravity ({\it e.g.} anomaly 
mediation), mediation by standard-model gauge interactions seems simpler 
to me, as it does not require a detailed understanding of Planck scale 
physics. 

Now, suppose that the sector responsible for dynamical supersymmetry 
breaking (DSB) is charged under standard-model gauge interactions.  Then 
it is possible that the gauginos obtain masses directly through their 
interaction with the DSB sector.  The squarks and sleptons in the SSM 
sector then obtain flavor universal masses through standard-model gauge 
interactions.  If this is the case, we do not need any other sector than 
the SSM and DSB sectors, which are in any case needed in any supersymmetric 
theory.  We just have to assume that the DSB sector is charged under 
standard-model gauge interactions, giving masses to the gauginos.  
What can be simpler than this? 

Constructing an explicit theory along the lines described above, however, 
is not an easy task.  Typically what happens is that if we want to 
make the DSB sector charged under standard-model gauge interactions, 
$SU(3)_C \times SU(2)_L \times U(1)_Y$, the gauge group $G$ of the DSB 
sector becomes large, making $SU(3)_C$ strongly asymptotically non-free, 
and the successful prediction for gauge coupling unification is 
lost~\cite{Dine:1993yw}.  In general, it is not at all easy to find 
an explicit gauge group and matter content for the DSB sector that does 
the required job and to construct a fully realistic theory.  One way out 
from this difficulty is to further separate the DSB sector from the SSM 
sector by introducing fields called messenger fields, which are charged 
both under standard-model gauge interactions and under interactions that 
mediate supersymmetry breaking from the DSB sector to the messenger 
fields~\cite{Dine:1994vc}.  This, however, loses a certain beauty that 
the original picture had. 

In this talk I want to present explicit theories in which the picture 
described above is realized in a simple way.  An important new 
ingredient is the correspondence between 4D gauge theories and their 
higher-dimensional dual gravitational descriptions, especially the AdS/CFT 
correspondence~\cite{Maldacena:1997re}.  This allows us to formulate our 
theories in higher dimensional spacetime, which does not require us to 
find the explicit gauge group and matter content for the DSB sector to 
construct the theories in a consistent effective field theory framework. 
In the construction, we require our theories to be fully realistic.  In 
particular, we require that the successful prediction for gauge coupling 
unification is preserved.  The theories are also free from problems of 
the simplest supersymmetric unified theories, such as the doublet-triplet 
splitting problem and the problem of overly rapid proton decay, and 
accommodate the successes of the conventional unification picture, such 
as the understanding of small neutrino masses by the see-saw mechanism. 
In much of the parameter space we find that the gauginos $\lambda$ and 
sfermions $\tilde{f}$ obtain masses of order $m_\lambda \sim (\alpha/4\pi) 
\Lambda$ and $m_{\tilde{f}}^2 \sim (\alpha/4\pi)^2 \Lambda^2$, respectively. 
This implies that the scale for supersymmetry breaking is rather low 
\begin{equation}
  \Lambda \approx 10\!\sim\!100~{\rm TeV}.
\end{equation}
This may be the lowest possible scale for supersymmetry breaking we can 
attain in realistic supersymmetric theories. 

This talk is mainly based on the works with Walter Goldberger, David 
Tucker-Smith, and Brock Tweedie, presented in Refs.~\cite{Goldberger:2002pc,%
Nomura:2003qb,Nomura:2004is,Nomura:2004it}.

\section{Supersymmetry in Warped Space}
\label{sec:susy-warped}

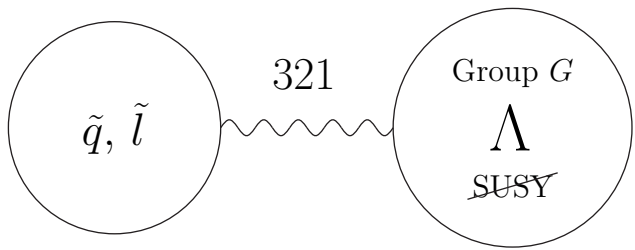
\begin{figure}[t]
\begin{center} 
\begin{picture}(350,95)(0,0)
  \Photon(140,50)(205,50){3}{5}  \Text(172,65)[b]{\Large 321}
  \CArc(100,50)(40,0,360)
  \Text(100,50)[]{\Large $\tilde{q},\,\tilde{l}$}
  \CArc(250,50)(45,0,360)
  \Text(251,71)[]{Group $G$} 
  \Text(250,50)[]{\huge $\Lambda$}
  \Text(251,28)[]{SUSY}  \Line(233,23)(267,33)
\end{picture}
\caption{The basic picture of our theories.}
\label{fig:structure}
\end{center}
\end{figure}

The theories we consider have the following basic structure.  We have a 
sector, the DSB sector, that breaks supersymmetry dynamically at a scale 
$\Lambda \approx 10\!\sim\!100~{\rm TeV}$.  We denote the gauge group of 
this sector as $G$.  This sector is also charged under the standard-model 
gauge group, $SU(3)_C \times SU(2)_L \times U(1)_Y$ (321).  Gaugino 
masses and flavor universal sfermion masses are then generated through 
standard-model gauge interactions.  This basic picture is depicted 
in Fig.~\ref{fig:structure}. 

In general, it is not easy to find an explicit calculable theory realizing 
the above basic structure, since it necessarily involves a sector that is 
strongly coupled at the scale $\Lambda \approx 10\!\sim\!100~{\rm TeV}$. 
Suppose now that the gauge coupling $\tilde{g}$ and the size (the number 
of ``colors'') $\tilde{N}$ of the group $G$ satisfy the following relation: 
$\tilde{g}^2 \tilde{N}/16\pi^2 \gg 1$ and $\tilde{N} \gg 1$.  In this case 
it is possible that the theory admits a dual higher-dimensional description 
that is weakly coupled, and allows explicit calculations of various 
quantities. 

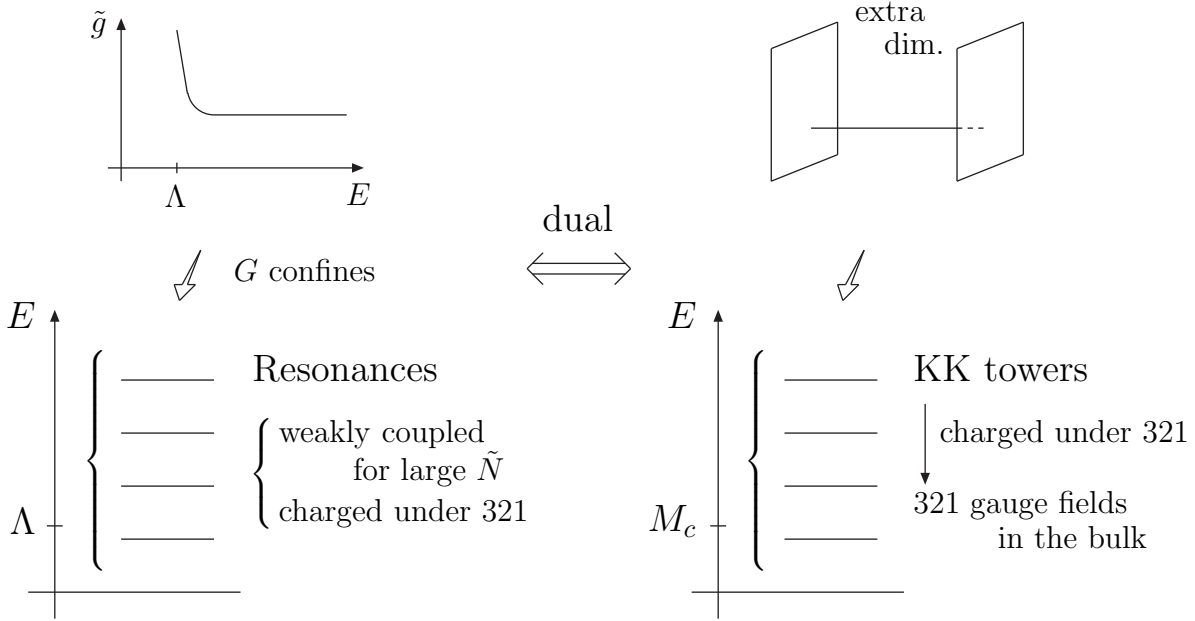
\begin{figure}[t]
\begin{center} 
\begin{picture}(470,235)(37,-5)
  \LongArrow(75,170)(170,170) \Text(170,165)[t]{$E$}
  \LongArrow(80,165)(80,225)  \Text(75,225)[r]{$\tilde{g}$}
  \Line(115,190)(165,190) \CArc(115,200)(10,190,270) \Line(105,198)(101,222)
  \Line(101,168)(101,172) \Text(101,164)[t]{$\Lambda$}
  \Line(110,139)(103,126) \Line(110,139)(105,126)
  \Line(100,126)(103,126) \Line(105,126)(108,126)
  \Line(100,126)(102,120) \Line(102,120)(108,126)
  \Text(123,132)[l]{$G$ confines}
  \Line(45,10)(125,10) \LongArrow(55,0)(55,115) \Text(48,115)[r]{\large $E$}
  \Line(52,35)(58,35)  \Text(48,37)[r]{\large $\Lambda$}
  \Line(80,30)(115,30) \Line(80,50)(115,50)
  \Line(80,70)(115,70) \Line(80,90)(115,90)
  \Text(80,60)[r]{$\left\{ \matrix{ \cr\cr\cr\cr\cr\cr } \right.$}
  \Text(130,95)[l]{\large Resonances}
  \Text(143,55)[r]{$\left\{ \matrix{ \cr\cr\cr } \right.$}
  \Text(140,70)[l]{weakly coupled}
  \Text(168,57)[l]{for large $\tilde{N}$}
  \Text(140,40)[l]{charged under 321}
  \Line(235,130)(270,130) \Line(235,134)(270,134)
  \Line(233,132)(239,138) \Line(233,132)(239,126)
  \Line(272,132)(266,138) \Line(272,132)(266,126)
  \Text(253,147)[b]{\large dual}
  \Line(325,165)(325,215) \Line(350,175)(350,225)
  \Line(325,165)(350,175) \Line(325,215)(350,225)
  \Line(395,165)(395,215) \Line(420,175)(420,225)
  \Line(395,165)(420,175) \Line(395,215)(420,225)
  \Line(340,185)(395,185) \DashLine(395,185)(405,185){2}
  \Text(358,230)[l]{extra}
  \Text(369,218)[l]{dim.}
  \Line(360,139)(353,126) \Line(360,139)(355,126)
  \Line(350,126)(353,126) \Line(355,126)(358,126)
  \Line(350,126)(352,120) \Line(352,120)(358,126)
  \Line(295,10)(375,10) \LongArrow(305,0)(305,115) \Text(298,115)[r]{\large $E$}
  \Line(302,35)(308,35)  \Text(298,37)[r]{\large $M_c$}
  \Line(330,30)(365,30) \Line(330,50)(365,50)
  \Line(330,70)(365,70) \Line(330,90)(365,90)
  \Text(330,60)[r]{$\left\{ \matrix{ \cr\cr\cr\cr\cr\cr } \right.$}
  \Text(380,95)[l]{\large KK towers}
  \LongArrow(383,81)(383,52) \Text(390,70)[l]{charged under 321}
  \Text(380,43)[l]{321 gauge fields}
  \Text(412,31)[l]{in the bulk}
\end{picture}
\caption{Schematic depiction of the correspondence between 
 the 4D and 5D theories.}
\label{fig:duality}
\end{center}
\end{figure}

The way this duality works is the following.  Let us first consider the 4D 
theory described in Fig.~\ref{fig:structure}.  In this theory the DSB sector 
exhibits non-trivial infrared dynamics at the scale $\Lambda$.  Besides 
dynamically breaking supersymmetry, this infrared dynamics produces a series 
of bound states, whose typical mass scale is the dynamical scale $\Lambda$. 
Since the DSB sector is charged under 321, these bound states are also charged 
under 321.  For $\tilde{N} \gg 1$, there are a large number of such bound 
states which are weakly coupled, as suggested by the analysis of large-$N$ 
QCD~\cite{'tHooft:1973jz}.  We now consider another theory formulated in 
higher dimensions, {\it e.g.} in 5D, and assume that the extra dimension is 
compactified with the characteristic mass scale for the Kaluza-Klein (KK) 
towers $M_c$.  Now, suppose that the spectrum of bound states obtained in 
the 4D theory of Fig.~\ref{fig:structure} and the KK spectrum of this 5D 
theory are exactly the same, $M_c \approx \Lambda$, and so are any physical 
quantities such as the scattering amplitudes among various states.  If this 
is the case, we can never distinguish the two theories experimentally, 
implying that the two theories just correspond to two different descriptions 
of the same physics.  This is the meaning of the duality, schematically 
depicted in Fig.~\ref{fig:duality}.  Because the bound states of the 4D 
theory are charged under 321, the KK towers of the 5D theory should also 
be charged under 321.  This implies that the 321 gauge fields must propagate 
in the 5D bulk in the ``dual'' 5D picture of the theory. 

What the ``dual'' 5D theory looks like more explicitly?  Let us now assume 
that the gauge coupling $\tilde{g}$ evolves very slowly above $\Lambda$, so 
that the $G$ sector is nearly conformal in a wide energy interval between 
$\Lambda$ and a high scale of order $M_X \simeq 10^{16}~{\rm GeV}$.  The 
AdS/CFT correspondence then implies that the 5D theory is formulated in 
anti-de~Sitter (AdS) space truncated by two branes, an ultraviolet (UV) 
brane and an infrared (IR) brane~\cite{Arkani-Hamed:2000ds}.  The metric 
of this spacetime is then given by
\begin{equation}
  ds^2 = e^{-2ky} \eta_{\mu\nu} dx^\mu dx^\nu + dy^2,
\label{eq:metric}
\end{equation}
where $y$ is the coordinate for the extra dimension and $k$ denotes the 
inverse curvature radius of the AdS space.  The two branes are located at 
$y=0$ (the UV brane) and $y=\pi R$ (the IR brane).  This is the spacetime 
considered in Ref.~\cite{Randall:1999ee}, in which the large hierarchy 
between the weak and the Planck scales are generated by the AdS warp factor.  
We here choose the scales such that the scales on the UV and IR branes 
are roughly the 4D Planck scale and the scale $\Lambda$, respectively: 
$k \sim M_5 \sim M_* \sim M_{\rm Pl}$ and $kR \sim 10$ (the 4D Planck scale 
is given by $M_{\rm Pl}^2 \simeq M_5^3/k$).  Here, $M_5$ is the 5D Planck 
scale, and $M_*$ the 5D cutoff scale, which is taken to be somewhat (typically 
a factor of a few) larger than $k$.  With this choice of scales, the 
characteristic mass scale for the KK towers is given by $\pi k e^{-\pi kR} 
\sim \Lambda \approx (10\!\sim\!100)~{\rm TeV}$. 

\begin{figure}[t]
\begin{center} 
\begin{picture}(250,192)(0,-20)
  \Text(60,165)[br]{``Planck'' brane}
  \Line(0,-15)(0,125) \Line(60,15)(60,155)
  \Line(0,-15)(60,15) \Line(0,125)(60,155)
  \Text(17,126)[l]{$Q,U,D$} \Text(25,113)[l]{$L,E$}
  \Line(30,20)(190,20) \DashLine(190,20)(220,20){3} \Vertex(30,20){1}
  \CArc(85,-325)(405,73,97)  \PhotonArc(85,-325)(405,73,97){3}{15}
  \CArc(85,449)(405,263,287) \PhotonArc(85,449)(405,263,287){3}{15}
  \Text(112,88)[b]{$\lambda$}
  \DashLine(36,77)(20,87){3} \Text(16,89)[r]{$\tilde{q}$}
  \DashLine(36,47)(20,37){3} \Text(16,36)[r]{$\tilde{q}$}
  \CArc(83,62)(50,162,198)  \Text(29,62)[r]{$q$}
  \Text(125,128)[b]{\large (321) gauge field}
  \Text(190,165)[bl]{``TeV'' brane}
  \Line(190,-15)(190,125) \Line(250,15)(250,155)
  \Line(190,-15)(250,15)  \Line(190,125)(250,155)
  \Text(221,118)[]{SUSY}  \Line(203,113)(237,123)
\end{picture}
\caption{The overall picture for the higher dimensional description of 
 the theory.}
\label{fig:overall}
\end{center}
\end{figure}
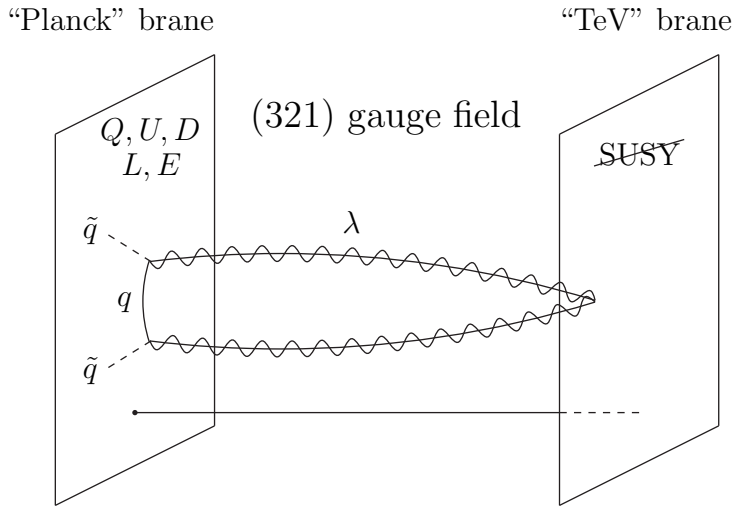

The theories described below~\cite{Goldberger:2002pc,Nomura:2003qb,%
Nomura:2004is,Nomura:2004it} are thus formulated in 5D supersymmetric warped 
space truncated by two branes.  The structure of Fig.~\ref{fig:structure} 
then corresponds to breaking supersymmetry on the IR brane (also called the 
TeV brane) and localizing quark and lepton superfields, $Q$, $U$, $D$, $L$ 
and $E$, to the UV brane (also called the Planck brane).  The standard-model 
gauge fields propagate in the bulk.  The overall picture is depicted 
in Fig.~\ref{fig:overall} (we can even see the similarity between the 
two pictures in Fig.~\ref{fig:structure} and Fig.~\ref{fig:overall}). 
Supersymmetry breaking on the TeV brane in this picture does not have to 
be suppressed --- it can be an $O(1)$ breaking when measured in terms of 
the 5D metric of Eq.~(\ref{eq:metric}).  Although supersymmetry breaking 
is directly transmitted to the 321 gauginos, the generated gaugino masses 
are of order TeV, because of the exponential warp factor.  The squark and 
slepton masses are also generated through 321 gauge loops, which are flavor 
universal and thus do not introduce the supersymmetric flavor problem. 
This setup was first considered in Ref.~\cite{Gherghetta:2000qt}.  We will 
see that in our theories this picture coexists with most of the successes 
of the conventional weak-scale supersymmetry paradigm. 

Here I want to emphasize that we should not take the view that our theory 
has solved the hierarchy problem {\it twice} by introducing both supersymmetry 
and a warped extra dimension.  Rather, the picture of a supersymmetric 
warped extra dimension arises if the DSB sector, which is necessarily present 
in any supersymmetric theory, satisfies certain conditions, {\it e.g.} 
$\tilde{g}^2 \tilde{N}/16\pi^2 \gg 1$ and $\tilde{N} \gg 1$.  A virtue of 
the higher dimensional construction is then that we do not need to know the 
gauge group or the matter content of the DSB sector explicitly.  In fact, 
once we have the picture of higher dimensional warped space and construct 
a theory on this space, we can forget about the ``original'' 4D picture for 
all practical purposes, although such a picture is useful for estimating 
various physical quantities and for obtaining insight on physical properties 
of the theory.  Although, strictly speaking, the presence of a higher 
dimensional theory does not necessarily guarantee the presence of the 
``corresponding 4D theory'', this need not concern us.  Our higher dimensional 
warped supersymmetric theory is a consistent effective field theory, with 
which we can calculate various physical quantities and compare with 
experiments --- the theory is even weakly coupled if the cutoff scale 
of the theory is sufficiently larger than the AdS curvature scale. 

In the next section we present a complete theory built on this picture, 
which accommodates the successes of the conventional weak-scale supersymmetry 
paradigm.  Alternative, related theories will be discussed in later sections.

\section{Warped Supersymmetric Grand Unification}
\label{sec:warped-sgut}

We start by recalling that our DSB sector is charged under the standard-model 
gauge group so that it contributes to the evolution of the 321 gauge couplings. 
On the other hand, the successful prediction for gauge coupling unification 
in the minimal supersymmetric standard model (MSSM) implies that any additional 
contributions to the evolution of the 321 gauge couplings beyond those from 
the MSSM gauge, matter and Higgs fields must be universal for $SU(3)_C$, 
$SU(2)_L$ and $U(1)_Y$.  In our context, this implies that the contribution 
from the DSB sector should be universal.  This is most naturally attained 
if the DSB sector possesses a global $SU(5)$ symmetry, of which the 321 gauge 
group of the standard model is a subgroup.  In the 5D picture, this corresponds 
to having a gauge group (at least) $SU(5)$ in the bulk, since the gauge 
symmetry in the 5D bulk corresponds to a global symmetry of the strong 
interacting sector in the 4D theory. 

The higher dimensional unified gauge group must be broken to the 321 subgroup, 
as the gauge invariance in our low-energy world is 321.  Because our theory 
looks higher dimensional, we can employ a higher-dimensional mechanism to break 
a gauge symmetry.  In particular, we can break 5D unified gauge invariance, 
taken as $SU(5)$ here, by imposing non-trivial boundary conditions on the 
fields at a boundary of the spacetime.  This way of breaking a unified 
gauge symmetry has many desirable features over the conventional Higgs 
mechanism; for example, it allows an elegant, simultaneous solution to 
the problems of doublet-triplet splitting, overly rapid proton decay and 
unwanted fermion mass relations~\cite{Hall:2001pg}.  Here we adopt this 
mechanism to construct our explicit theory.

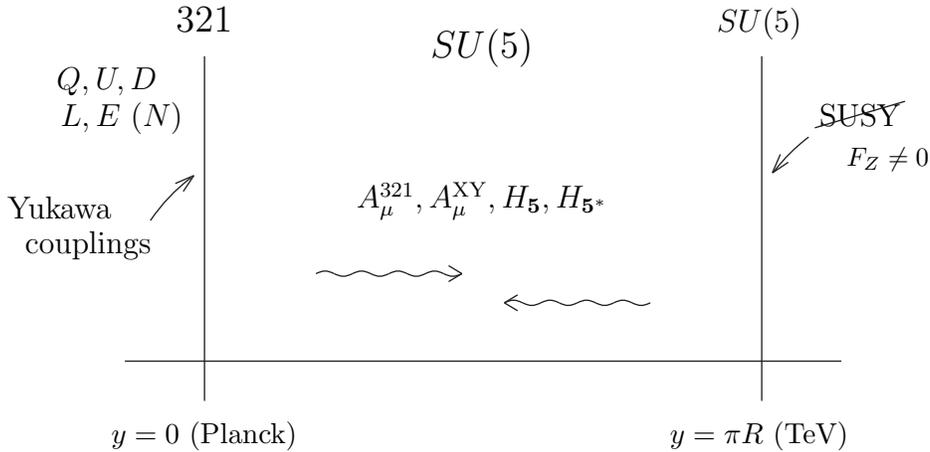
\begin{figure}[t]
\begin{center} 
\begin{picture}(270,175)(0,-33)
  \Line(0,5)(270,5) \Text(135,120)[b]{\large $SU(5)$}
  \Line(30,-10)(30,120) \Text(30,130)[b]{\large $321$} 
  \Text(30,-18)[t]{\small $y=0$ (Planck)}
  \Text(12,111)[r]{$Q,U,D$} \Text(22,97)[r]{$L,E$ $(N)$}
  \CArc(66,20)(68,127,146) \Line(26,75)(22,69) \Line(26,75)(19,72)
  \Text(-5,63)[r]{Yukawa} \Text(10,49)[r]{couplings}
  \Line(240,-10)(240,120) \Text(240,130)[b]{$SU(5)$}
  \Text(240,-18)[t]{\small $y=\pi R$ (TeV)}
  \CArc(298,36)(67,127,143) \Line(244,76)(247,83) \Line(244,76)(250,80)
  \Text(278,98)[]{SUSY}  \Line(260,93)(294,103)
  \Text(273,86)[tl]{\footnotesize $F_Z \neq 0$}
  \Text(135,63)[b]{$A_\mu^{321}, A_\mu^{\rm XY}, H_{\bf 5}, H_{{\bf 5}^*}$}
  \Photon(72,38)(127,38){1}{4}  \Line(127,38)(122,41) \Line(127,38)(122,35)
  \Photon(143,27)(198,27){1}{4} \Line(143,27)(148,30) \Line(143,27)(148,24)
\end{picture}
\caption{The overall picture of the theory.}
\label{fig:theory}
\end{center}
\end{figure}

The structure of our minimal warped supersymmetric unified theory is then 
given as follows~\cite{Goldberger:2002pc}.  The theory is formulated in 5D 
warped space with the metric given by Eq.~(\ref{eq:metric}) and the extra 
dimension compactified on an interval $0 \leq y \leq \pi R$.  The bulk gauge 
group is taken to be $SU(5)$, which is broken by boundary conditions at $y=0$ 
(the Planck brane).  The two Higgs hypermultiplets, which are ${\bf 5}$ 
and ${\bf 5}^*$ representations under $SU(5)$, are introduced in the bulk. 
Depending on the values of the bulk masses for the Higgs multiplets, 
which are conveniently parameterized by two dimensionless numbers $c_H$ and 
$c_{\bar{H}}$ (for notation see~\cite{Goldberger:2002pc}), the wavefunctions 
for the zero modes arising from these multiplets can have varying shapes. 
The matter fields are localized to the Planck brane --- they could either 
be located on the Planck brane or be introduced in the bulk but with the 
zero modes strongly localized to the Planck brane by bulk mass terms. 
The Yukawa couplings are also located on the Planck brane.  Supersymmetry 
is broken on the TeV brane by a vacuum expectation value (VEV) for the 
auxiliary field of a chiral superfield $Z$.  The overall picture of the 
theory is depicted in Fig.~\ref{fig:theory}.

The boundary conditions for various fields are given more explicitly 
in Table~\ref{table:bc}.  Here we represent the 5D gauge multiplet 
${\cal V} \equiv \{V, \Sigma\}$ in terms of a 4D $N=1$ vector 
superfield $V$ and a chiral superfield $\Sigma$.  The subscripts 
321 and XY represent $321$ and $SU(5)/321$ components, respectively. 
A bulk hypermultiplet is represented by two 4D $N=1$ chiral superfields 
$\Phi$ and $\Phi^c$.  Therefore, the two Higgs hypermultiplets are 
denoted as $\{ H, H^c \}$ and $\{ \bar{H}, \bar{H}^c \}$, with the 
subscripts $D$ and $T$ representing the doublet and triplet components, 
respectively.  The boundary conditions are written in the language of 
orbifolding procedures.  In the table we also give the boundary conditions 
for bulk matter fields.  For bulk matter, we need two hypermultiplets 
$\{ T, T^c \} + \{ T', T'^c \}$ in the ${\bf 10}$ representation and 
two hypermultiplets $\{ F, F^c \} + \{ F', F'^c \}$ in the ${\bf 5}^*$ 
representation to complete a single generation.  The subscripts for these 
fields denote the irreducible components under the 321 decompositions.%
\footnote{Our theory can also be viewed as one with $SU(5)$ broken by a large 
VEV of a Higgs field on the Planck brane (see {\it e.g.}~\cite{Nomura:2001mf}), 
although in that case an understanding of the doublet-triplet splitting 
should be attributed to unknown cutoff scale physics.}
\begin{table}[t]
\begin{center}
\begin{tabular}{|c|c|c|}
\hline
 $(p,p')$  &  gauge and Higgs fields  & 
    bulk matter fields \\ \hline
 $(+,+)$  & $V_{321}$,         $H_D$,   $\bar{H}_D$   & 
    $T_{U,E}$, $T'_Q$,     $F_D$, $F'_L$      \\ 
 $(-,-)$  & $\Sigma_{321}$,    $H^c_D$, $\bar{H}^c_D$ & 
    $T^c_{U,E}$, $T'^c_Q$, $F^c_D$, $F'^c_L$  \\ 
 $(-,+)$  & $V_{\rm XY}$,      $H_T$,   $\bar{H}_T$   & 
    $T_Q$, $T'_{U,E}$,     $F_L$, $F'_D$      \\ 
 $(+,-)$  & $\Sigma_{\rm XY}$, $H^c_T$, $\bar{H}^c_T$ & 
    $T^c_Q$, $T'^c_{U,E}$, $F^c_L$, $F'^c_D$  \\ 
\hline
\end{tabular}
\caption{The boundary conditions for the bulk fields. The fields written 
 in the $(p,p')$ column, $\varphi$, obey the boundary condition 
 $\varphi(-y) = p\, \varphi(y)$ and $\varphi(-y') = p'\, \varphi(y')$ when 
 we construct our space, $0 \leq y \leq 2\pi$, by the orbifolding procedure. 
 Here, $y' \equiv y-\pi R$.}
\label{table:bc}
\end{center}
\end{table}

\begin{figure}[t]
\begin{center} 
\begin{picture}(385,165)(-10,-18)
  \Line(5,0)(355,0) \LongArrow(15,-10)(15,135) 
  \Text(8,135)[r]{mass} \Text(8,75)[r]{$\sim {\rm TeV}$} 
  \Text(45,133)[b]{$V_{321}$}
  \Line(30,0)(60,0)     \Vertex(45,0){3}
  \Line(30,45)(60,45)   \Vertex(45,45){3}
  \Line(30,105)(60,105) \Vertex(45,105){3}
  \Text(85,133)[b]{$\Sigma_{321}$}
  \Line(70,45)(100,45)   \Vertex(85,45){3}
  \Line(70,105)(100,105) \Vertex(85,105){3}
  \Text(125,133)[b]{$V_{\rm XY}$}
  \Line(110,45)(140,45)   \Vertex(125,45){3}
  \Line(110,105)(140,105) \Vertex(125,105){3}
  \Text(165,133)[b]{$\Sigma_{\rm XY}$}
  \Line(150,45)(180,45)   \Vertex(165,45){3}
  \Line(150,105)(180,105) \Vertex(165,105){3}
  \Text(205,133)[b]{$H_{\!D},\!\bar{H}_{\!D}$}
  \Line(190,0)(220,0)   \Vertex(205,0){3}
  \Line(190,80)(220,80) \Vertex(205,80){3} 
  \Text(245,133)[b]{$H_{\!D}^c,\!\bar{H}_{\!D}^c$}
  \Line(230,80)(260,80) \Vertex(245,80){3} 
  \Text(285,133)[b]{$H_{\!T},\!\bar{H}_{\!T}$}
  \Line(270,80)(300,80) \Vertex(285,80){3} 
  \Text(325,133)[b]{$H_{\!T}^c,\!\bar{H}_{\!T}^c$}
  \Line(310,80)(340,80) \Vertex(325,80){3} 
\end{picture}
\caption{The schematic picture for the lowest KK spectrum.}
\label{fig:KK}
\end{center}
\end{figure}
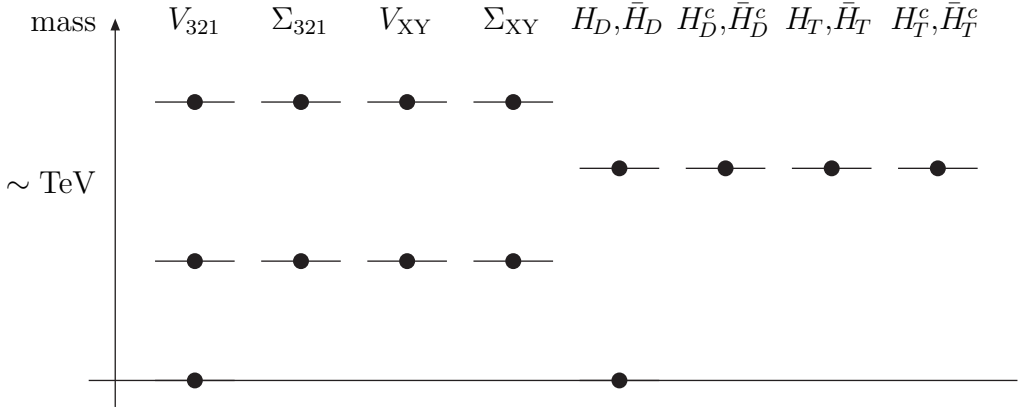

The spectrum of the theory is obtained by KK decomposing the fields.  
We then find that the zero-mode sector contains only the MSSM fields: 
the 321 gauge field, $V_{321}$, two Higgs doublets, $H_D$ and $\bar{H}_D$, 
and three generations of matter fields $Q$, $U$, $D$, $L$ and $E$.  The 
characteristic mass scale for the KK towers is of order TeV, $M_c \equiv 
\pi k e^{-\pi kR}$.  They are almost $N=2$ supersymmetric and $SU(5)$ 
symmetric.  For example, for the gauge sector, each KK level contains 
$V_{321}$, $\Sigma_{321}$, $V_{\rm XY}$ and $\Sigma_{\rm XY}$.  The 
spectrum for a characteristic case (for characteristic values of $c_H$ and 
$c_{\bar{H}}$) is depicted schematically in Fig.~\ref{fig:KK}.  Because the 
spectrum at the TeV scale has a radical departure from that of the MSSM, one 
might wonder to what extent the successes of the conventional supersymmetric 
desert scenario are preserved.  Below we will see that our theory preserves 
most of the successes of the conventional desert scenario and, moreover, 
is free from the problems which the minimal supersymmetric unified theory 
suffers from. 

\subsection{Gauge coupling unification}
\label{subsec:gcu}

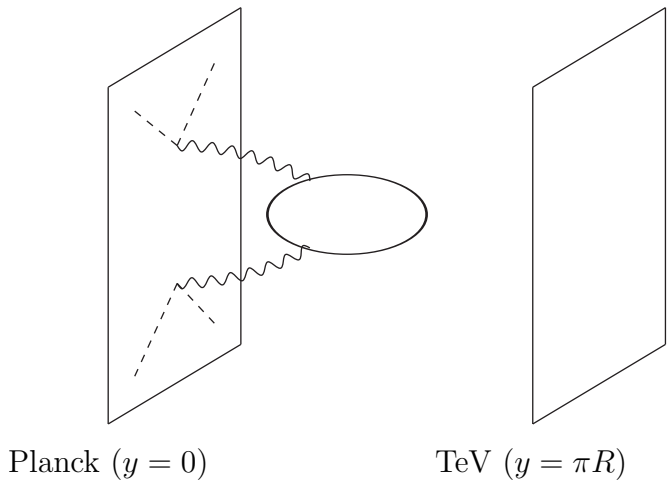
\begin{figure}[t]
\begin{center} 
\begin{picture}(210,185)(0,-35)
  \Text(0,-16)[t]{Planck ($y=0$)}
  \Line(0,-7)(0,120) \Line(50,23)(50,150)
  \Line(0,-7)(50,23) \Line(0,120)(50,150)
  \DashLine(26,46)(10,11){3}  \DashLine(26,46)(40,31){3}
  \DashLine(26,98)(40,129){3} \DashLine(26,98)(10,111){3} 
  \Oval(90,72)(15,30)(0)
  \PhotonArc(26,-2)(100,60,90){2}{7}
  \PhotonArc(26,146)(100,270,300){2}{7}
  \Text(160,-16)[t]{TeV ($y=\pi R$)}
  \Line(160,-7)(160,120) \Line(210,23)(210,150)
  \Line(160,-7)(210,23)  \Line(160,120)(210,150)
\end{picture}
\caption{A diagram giving the gauge coupling evolution in warped space.}
\label{fig:diag}
\end{center}
\end{figure}

As is suggested from the correspondence between the 4D and 5D pictures, the 
evolution of the gauge couplings in our theory is logarithmic.  The fact 
that the gauge couplings for bulk gauge fields evolve logarithmically in 
warped space was first noticed in Ref.~\cite{Pomarol:2000hp}, in which the 
successful prediction was also anticipated based on a heuristic argument. 
There have been some debates on whether theories in warped space actually 
allow calculations of gauge coupling unification; in particular, 
whether threshold corrections at an IR scale are under control (see 
{\it e.g.}~\cite{Arkani-Hamed:2000ds}).  Subsequent theoretical works, 
however, have clarified that these corrections are in fact under control, 
and that theories on warped space retain calculability~\cite{Randall:2001gc}. 
For an observer sitting at $y=y_*$, physics is essentially four dimensional 
up to an energy $E \sim k e^{-k y_*}$, so for the Planck-brane observer 
physics is four dimensional all the way up to $k \sim M_{\rm Pl}$.  Now, 
the gauge couplings are measured, for example, by scattering two quarks 
--- a process that occurs on the Planck brane.  The evolution of the gauge 
couplings are then given by calculating diagrams as given in Fig~\ref{fig:diag} 
and summing up logarithms arising from them.  At energies higher than the 
TeV scale $E \gg k' \sim {\rm TeV}$, the gauge propagator in the bulk cannot 
probe the region close to the TeV brane, as the propagation of a gauge field 
from $y=0$ to $y=\pi R$ receives a large suppression, $\propto \exp(-E/k')$, 
for $E \simgt k'$.  In warped space all the KK modes are strongly localized 
to the TeV brane except for a single mode, which is often the zero mode. 
This implies that the contribution to the evolution of the gauge couplings 
at $E \simgt {\rm TeV}$ is dominated by the single mode and thus is 
logarithmic.  For the case of a non-Abelian gauge field, the situation is 
somewhat more complicated due to the mass mixing between the different 
modes, but the essential physics is still the same and the evolution is 
still dominated by ``a single mode'' and is four dimensional. 

In Ref.~\cite{Goldberger:2002pc} we showed that the successful prediction 
for gauge coupling unification (the same prediction as the MSSM) is 
automatically obtained if the following two conditions are obeyed: 
\begin{itemize}
\item The bulk gauge group is $SU(5)$ (or a larger group containing $SU(5)$) 
that is broken on the Planck brane (at the scale $k$ or larger).
\item The bulk mass parameters for the matter and Higgs fields are all 
larger than or equal to $1/2$: $c_{\rm matter}, c_{\rm Higgs} \geq 1/2$. 
This implies that zero modes for these fields have wavefunctions either 
conformally flat or localized to the Planck brane.
\end{itemize}
If the breaking scale of $SU(5)$ is somewhat larger than $k$, such as the 
case of boundary condition breaking (breaking by the Planck-brane boundary 
conditions corresponds to breaking $SU(5)$ at a scale much larger than $k$), 
tree-level operators on the Planck brane could potentially give incalculable 
non-universal corrections.  These corrections, however, are naturally 
suppressed if the volume of the bulk is large, which is necessarily the 
case in warped space theories explaining the hierarchy between the Planck 
and the TeV scales.  In our theory matter is localized to the Planck brane, 
corresponding to $c_{\rm Matter} \gg 1/2$.  The only remaining condition 
is then that the two Higgs multiplets must have mass parameters larger 
than or equal to $1/2$: $c_H, c_{\bar{H}} \geq 1/2$.  Under this condition, 
we find that the prediction for the low-energy 321 gauge couplings in our 
theory is given by
\begin{equation}
  \pmatrix{1/g_1^2 \cr 1/g_2^2 \cr 1/g_3^2}_{\! \mu = M_Z}
  \simeq \:\: (SU(5)\,\,\, {\rm symmetric}) 
    + \frac{1}{8 \pi^2} \pmatrix{33/5 \cr 1 \cr -3} 
      \ln\left(\frac{k}{M_Z}\right),
\label{eq:gc-low}
\end{equation}
which is identical to the MSSM prediction, with the AdS curvature 
$k$ identified as the conventional unification scale $M_X \simeq 
10^{16}~{\rm GeV}$.  This determines the scales in the theory to 
be $k \simeq 10^{16-17}~{\rm GeV}$ and $M_* \simeq M_5 \simeq 
10^{17-18}~{\rm GeV}$.  It is fortunate that we obtain these numbers, 
as we obtain roughly the correct size for the 4D Planck scale 
$M_{\rm Pl} \simeq (M_5^3/k)^{1/2}$ without introducing a new scale. 

It is useful to consider gauge coupling unification in the 4D picture. 
In the 4D picture our theory appears as follows~\cite{Goldberger:2002pc}. 
We have the DSB sector with the gauge group $G$, whose coupling $\tilde{g}$ 
evolves very slowly over a wide energy interval between $k \approx M_X 
\sim 10^{16}~{\rm GeV}$ and $k' \approx \Lambda \sim {\rm TeV}$.  The value 
of the coupling is $\tilde{g} \simeq 4\pi$ in this energy interval, and the 
size of the gauge group $\tilde{N}$ is sufficiently larger than $1$ so that 
$\tilde{g}^2\tilde{N}/16\pi^2 \gg 1$.  The DSB sector possesses a global 
$SU(5)$ symmetry, of which the 321 subgroup is gauged and identified as 
the standard-model gauge group.  Quark, lepton and two Higgs-doublet 
superfields are introduced as elementary fields, which interact with the 
DSB sector through 321 gauge interactions.  The Higgs fields may also 
have direct interactions with the DSB sector through couplings of the form 
${\cal L} \sim H {\cal O}_H + \bar{H} {\cal O}_{\bar{H}}$, where ${\cal O}_H$ 
and ${\cal O}_{\bar{H}}$ are operators of the DSB sector.  The strengths of 
these couplings in the IR depend on the parameters $c_H$ and $c_{\bar{H}}$ 
in the 5D picture.  Once supersymmetry is broken at the scale $\Lambda$ by 
the non-trivial IR dynamics of $G$, the 321 gauginos, squarks and sleptons 
(and the Higgs fields) receive masses through 321 gauge interactions. 

\begin{figure}[t]
\begin{center}
\begin{picture}(300,148)(-15,-23)
  \DashCArc(254,120)(220,180,193){3} 
  \DashCArc(50,78)(12,180,270){3} 
  \Line(50,66)(230,66)  \Text(238,66)[l]{$\tilde{g}$}
  \CArc(-2,657)(650,271.1,291)  \Text(10,5)[r]{\small $g_1$}
  \CArc(-3,815)(800,271,287)    \Text(10,15)[r]{\small $g_2$}
  \CArc(1,1126)(1100,270.6,282) \Text(10,27)[r]{\small $g_3$}
  \DashLine(-10,50)(260,50){2} \Text(-15,50)[r]{$1$}
  \DashLine(50,0)(50,120){2} \Text(50,-5)[t]{$\Lambda \sim k'$}
  \DashLine(230,66)(230,0){2} \Vertex(230,66){2} \Text(230,-5)[t]{$k$}
  \LongArrow(-10,-8)(-10,120) \Text(-15,120)[r]{$\frac{g^2 N}{16\pi^2}$}
  \LongArrow(-18,0)(260,0) \Text(259,-6)[t]{$E$}
\end{picture}
\caption{Schematic description for the evolution of the gauge couplings 
 in our theory.} 
\label{fig:couplings}
\end{center}
\end{figure}
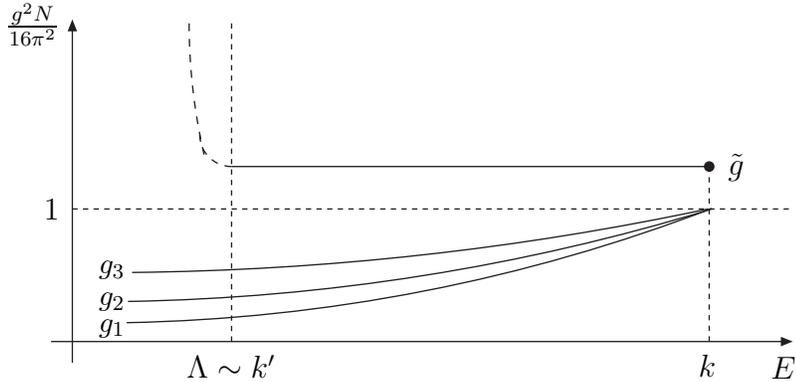

Since the DSB sector is charged under 321, the evolution of the 321 
gauge couplings receives a contribution from this sector as well as that 
from the elementary states.  At low energies $q \sim {\rm TeV}$, the 321 
gauge couplings are thus given by
\begin{equation}
  \frac{1}{g_a^2(q)} = \frac{1}{g_a^2(k)} 
      + {b_{\rm DSB} \over 8\pi^2} \ln\left({k \over q}\right)
      + {b_a \over 8\pi^2} \ln\left({k \over q}\right),
\label{eq:gc-4D}
\end{equation}
where $b_{\rm DSB}$ $(>0)$ represents the contribution from the DSB sector, 
which is universal due to the global $SU(5)$ symmetry, and $b_a$ the 
contribution from the elementary states: $(b_1, b_2, b_3) = (33/5, 1, -3)$. 
Now, in the 4D theory dual to the 5D theory with boundary condition 
$SU(5)$ breaking, the value of $b_{\rm DSB}$ is given such that the UV 
values of the 321 gauge couplings are strong, $g_a(k) \sim 4\pi$, 
in which case the first and second terms of the right-hand-side of 
Eq.~(\ref{eq:gc-4D}) are of $O(1/16\pi^2)$ and $O(1)$, respectively 
(the actual value is $b_{\rm DSB} \approx 5$).  It is then clear that the 
contributions from these terms are approximately $SU(5)$ symmetric, so 
that the differences of the three couplings at low energies are essentially 
given by the last term.  This gives the same prediction for gauge coupling 
unification as that of the MSSM.  The schematic picture for the evolution 
of the gauge couplings are given in Fig.~\ref{fig:couplings}.  It is 
interesting to note that in our theory the hierarchy between the Planck 
and the weak scales are generated by
\begin{equation}
  |\tilde{b}| \ll 1, \qquad \tilde{g} \sim 4\pi,
\end{equation}
where $\tilde{b}$ is the beta-function coefficient for the evolution 
of $\tilde{g}$, while in the conventional picture it is generated by 
$|\tilde{b}| \sim 1$ and $\tilde{g} \ll 4\pi$ (see Eq.~(\ref{eq:dim-trans})).

\subsection{Proton decay}
\label{subsec:p-decay}

Since the spectrum of the theory contains the XY gauge bosons and colored 
Higgs triplets at the TeV scale, one might worry that proton decay occurs 
at a disastrous rate in our theory.  However, this should not be the case 
if quark and lepton fields are localized to the Planck brane.  The scales 
on this brane do not receive a suppression by a warp factor under the 
dimensional reduction, so that any proton decay operator generated by 
integrating out baryon-number violating physics should be suppressed by 
a large mass scale of order $k$ in the low-energy 4D theory.  This must 
always be the case as long as the scale of $SU(5)$ breaking is at or 
larger than $k$ (the boundary condition breaking can be regarded as 
the breaking at the scale much higher than $k$).

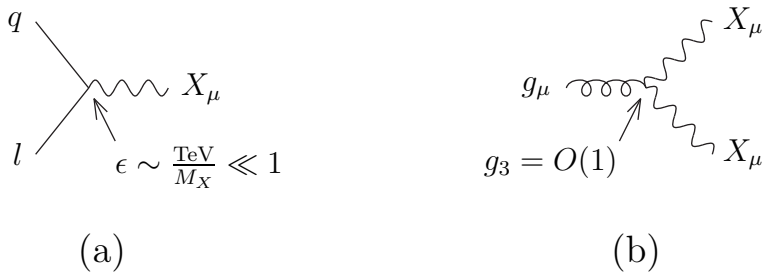
\begin{figure}[t]
\begin{center} 
\begin{picture}(255,100)(0,-50)
  \Text(25,-30)[t]{\large (a)}
  \Line(0,0)(20,25)  \Text(-5,0)[r]{$l$}
  \Line(0,50)(20,25) \Text(-5,50)[r]{$q$}
  \Photon(20,25)(50,25){3}{3} \Text(55,25)[l]{$X_\mu$}
  \Line(22,21)(30,5) \Line(22,21)(23,13) \Line(22,21)(27,16)
  \Text(30,3)[tl]{$\epsilon \sim \frac{\rm TeV}{M_X} \ll 1$}
  \Text(227,-30)[t]{\large (b)}
  \Gluon(200,25)(230,25){3}{3.5} \Text(195,25)[r]{$g_\mu$}
  \Photon(230,25)(255,0){3}{3.5}  \Text(260,0)[l]{$X_\mu$}
  \Photon(230,25)(255,50){3}{3.5} \Text(260,50)[l]{$X_\mu$}
  \Line(228,21)(220,5) \Line(228,21)(227,13) \Line(228,21)(223,16)
  \Text(220,3)[tr]{$g_3 = O(1)$}
\end{picture}
\caption{Vertices for the XY gauge bosons.}
\label{fig:pd}
\end{center}
\end{figure}

One might still wonder how the suppression of proton decay is explicitly 
realized in the KK decomposed 4D picture.  To see this, note that the 
wavefunctions of the XY gauge bosons and the colored Higgs triplets 
in our theory are strongly localized to the TeV brane (this is exactly 
the reason why these states have TeV-scale masses --- the masses for 
these states arise from the curvatures of the wavefunctions, which are 
localized to the TeV brane, so that they receive strong suppressions 
from a large warp factor).  Therefore, the wavefunction overlaps of 
the XY-gauge or colored-Higgs states to the quark and lepton fields 
are exponentially small of $O({\rm TeV}/M_X)$.%
\footnote{In the theory with boundary condition $SU(5)$ breaking, the 
minimal couplings of the XY gauge bosons to quarks and leptons vanish, 
but the couplings of the XY-gauge or colored-Higgs fields to matter 
can still arise in 4D from the 5D couplings that involve a derivative 
with respect to the fifth coordinate.}
This leads to tiny couplings for the baryon-number-violating 
vertices such as the (quark)-(lepton)-(XY gauge bosons) vertex 
(see Fig.~\ref{fig:pd}a), and thus suppresses any proton decay 
process to the level of conventional unified theories, given by 
${\cal L}_{\rm eff} \sim qqql/M_X^2$ or $qq\tilde{q}\tilde{l}/M_X$. 
Incidentally, the fact that the coupling in Fig.~\ref{fig:pd}a is tiny 
does not preclude the possibility of producing the XY gauge states at 
colliders, as they can be produced through the coupling to the gluon, 
which is the QCD coupling and $O(1)$ (see Fig.~\ref{fig:pd}b).  This 
coupling, of course, does not lead to proton decay because it conserves 
baryon number. 

In supersymmetric theories, it is in general not sufficient to ensure 
that proton decay operators are suppressed by the unified mass scale 
$M_X \simeq 10^{16}~{\rm GeV}$, because the dimension-five operators 
such as $W_{\rm eff} \sim QQQL/M_X$ could cause proton decay at a level 
contradicting to the experiments.  In our theory, however, dimension-five 
proton decay operators are simply absent because of a $U(1)_R$ symmetry 
the original 5D theory possesses.  The charges of various 4D $N=1$ 
superfields under $U(1)_R$ are given in Table~\ref{table:U1R}.  This 
symmetry clearly forbids the dimension-five proton decay operators 
$W \sim QQQL/M_X$ and $UUDE/M_X$, as well as other phenomenologically 
dangerous operators such as $W \sim M_X H \bar{H}$.  Diagrammatically, the 
absence of dimension-five proton decay due to the Higgs triplet exchange 
is understood by the structure of the mass terms for the Higgs triplets: 
$W \sim H_T H_T^c + \bar{H}_T \bar{H}_T^c$, which is different from that 
in the minimal supersymmetric grand unified theory $W \sim H_T \bar{H}_T$. 
After supersymmetry is broken, $U(1)_R$ is broken to its $Z_{2,R}$ 
subgroup, which is exactly $R$ parity.  Thus, dangerous dimension-four 
proton decay operators are strictly forbidden, and the supersymmetric 
mass term for the Higgs doublets of order the weak scale can be generated. 
The breaking $U(1)_R \rightarrow Z_{2,R}$ does not reintroduce 
dimension-five proton decay at a dangerous level.
\begin{table}[t]
\begin{center}
\begin{tabular}{|c|cc|cccc|cccccc|}  \hline 
  & $V$ & $\Sigma$ & $H$ & $H^c$ & $\bar{H}$ & $\bar{H}^c$ 
  & $T$ & $T^c$ & $F$ & $F^c$ & $N$ & $N^c$\\ \hline
  $U(1)_R$ & 0 & 0 & 0 & 2 & 0 & 2 & 1 & 1 & 1 & 1 & 1 & 1 \\ \hline
\end{tabular}
\end{center}
\caption{$U(1)_R$ charges for 4D vector and chiral superfields. The charges 
 of the primed matter fields, $T', T'^c, F', F'^c, N'$ and $N'^c$, are 
 the same as the non-primed fields.}
\label{table:U1R}
\end{table}

We finally mention that unwanted fermion mass relations such as $m_s/m_d 
= m_\mu/m_e$ does not arise in our theory.  This is because the Yukawa 
couplings are located on the Planck brane, on which the gauge group is 
reduced to 321.  An extension of the theory leading to successful 
$b/\tau$ unification will be discussed in section~\ref{subsec:SO10}.

\subsection{Other issues}
\label{subsec:other}

We have seen that in our theory two scales coexist in an intriguing way. 
On one hand, we have an ``extra dimension'' at the TeV scale, which is 
characterized by the appearance of the KK towers (including those for the 
grand-unified states) at the TeV scale.  The scale of supersymmetry breaking 
is also naturally set by this scale.  On the other hand, we have a very high 
scale $k \approx M_X \simeq 10^{16}~{\rm GeV}$ at which the 321 gauge 
couplings ``unify''.  This is also the scale that suppresses effective 
proton decay operators.  The reason why the scales for gauge coupling 
unification and proton decay are high is that we have broken $SU(5)$ at 
the Planck brane, where the warp factor is $1$ and does not give any 
suppression of the scales.  We can now go further in using these two 
coexisting mass scales.  For example, if we introduce right-handed neutrinos 
$N$ and introduce the Majorana masses and Yukawa couplings on the Planck 
brane, {\it i.e.} $\delta(y) \int d^2 \theta ((M_N/2)N^2+y_\nu LHN)$, we 
obtain small observed neutrino masses naturally through the conventional 
see-saw mechanism, because the Majorana masses, $M_N$, do not receive 
any suppressions from the warp factor. 

In fact, the correspondence between the 4D and 5D pictures suggests that 
our theory can be interpreted as a purely 4D theory, in which physics between 
the weak and unification scales is simply 4D $N=1$ supersymmetric field theory. 
This implies, for example, that the cosmological evolution in the early 
universe is purely four dimensional in our theory.  It is interesting to 
note that the theory is free from dangerous relics such as the gravitino 
and moduli.  Because the supersymmetry-breaking scale is very low, $\Lambda 
\approx (10\!\sim\!100)~{\rm TeV}$, we expect that the gravitino (and 
moduli, if any) is very light 
\begin{equation}
  m_{3/2} \simeq \frac{\Lambda^2}{M_{\rm Pl}} 
    \approx (0.1\!\sim\!10)~{\rm eV},
\end{equation}
in our theory.  Such a light gravitino does not produce the ``gravitino 
problem'', as its thermal relic abundance is small.  We also note that in 
warped theories, the radion does not cause any cosmological problem, as its 
mass and interaction strengths are both dictated by the TeV scale so that 
it decays before the big-bang nucleosynthesis.  Dark matter in our theory 
may come from conventional candidates such as axion, or may arise from 
a particle localized on the TeV brane, which has naturally TeV-scale mass 
and interactions and whose decay is protected by some discrete symmetry.%
\footnote{An alternative possibility is that there is an additional source 
of supersymmetry breaking on the Planck brane (such a breaking may in fact 
help to stabilize the radius of the extra dimension).  As long as the scale 
$\Lambda'$ of this additional breaking is $\Lambda' \simlt 10^9~{\rm GeV}$, 
this does not reintroduce the supersymmetric flavor problem, even in the 
presence of generic interactions between the supersymmetry-breaking field 
on the Planck brane and the quark-lepton supermultiplets.  This allows 
a wide range of possibilities for the gravitino mass, $0.1~{\rm eV} \simlt 
m_{3/2} \simlt 1~{\rm GeV}$, some of which is consistent with the scenario 
that the gravitino is the super-WIMP dark matter of the universe.}

\section{Phenomenology}
\label{sec:pheno}

The phenomenology of our theory is, naturally, quite rich, as it predicts 
a plethora of new particles at the TeV scale --- superparticles and $SU(5)$ 
states as well as their KK towers (superparticles even form $N=2$ multiplets 
at higher KK level).  In this section we study various aspects of the 
phenomenology associated with these particles.

\subsection{Spectrum}
\label{subsec:spectrum}

\begin{figure}[t]
  \center{\includegraphics[width=.7\textwidth]{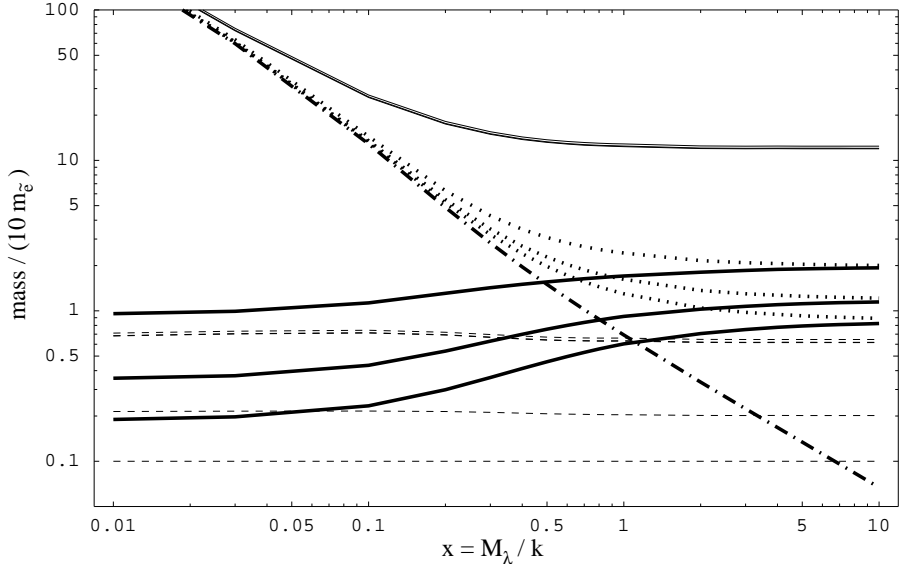}}
\caption{Masses of the MSSM scalars (dashed, with $m_{\tilde{q}}$, 
 $m_{\tilde{u}}$, and $m_{\tilde{d}}$ closely spaced and $m_{\tilde{l}}$ 
 and $m_{\tilde{e}}$ below), MSSM gauginos (thick solid), XY gauginos 
 (dot-dashed), 321 gaugino KK modes (dotted), and XY and 321 KK gauge 
 bosons (thin solid, nearly degenerate and most massive). The masses 
 are given in units of $10 \, m_{\tilde{e}}$.}
\label{fig:loglog}
\end{figure}

In the minimal warped unified theory described in section~\ref{sec:warped-sgut}, 
the spectrum of the TeV states are determined essentially by only two 
free parameters (up to parameters in the Higgs sector).  This is because 
supersymmetry breaking occurs on the TeV brane, on which the gauge group 
is effectively $SU(5)$: supersymmetry breaking is transmitted to the SSM 
sector through the operator
\begin{equation}
  {\cal L} = \delta(y-\pi R)\int d^2\theta 
      \frac{\zeta Z}{M_*}{\cal W}_a^\alpha {\cal W}_{a,\alpha}
    \:\: \rightarrow \:\: 
      \delta(y-\pi R) M_\lambda\, \lambda_a^\alpha \lambda_{a,\alpha}.
\label{eq:gaugino-TeV}
\end{equation}
which has only a single coupling for $a=SU(3)_C, SU(2)_L$ and $U(1)_Y$ 
($M_\lambda$ does not depend on $a$).  This allows us to calculate soft 
supersymmetry breaking parameters in terms of two parameters $x \equiv 
M_\lambda/k$ and $k' = ke^{-\pi kR}$~\cite{Nomura:2003qb}.  The result 
of this calculation is shown in Fig.~\ref{fig:loglog}.  In the figure, 
we have normalized all the masses in units of $10 \, m_{\tilde{e}}$.

In the supersymmetric limit ($x=0$), the spectrum consists of the MSSM 
states, which are massless, and the KK states, which are $SU(5)$ symmetric 
and $N=2$ supersymmetric.  Once supersymmetry is broken ($x \neq 0$), the 
321 gauginos, squarks and sleptons obtain masses.  In the meantime, the 
masses for the KK states are also shifted; in particular, one of the 
two 321 gauginos at each level becomes lighter and the other heavier, 
and similarly for the XY gauginos.  In the limit of large supersymmetry 
breaking ($x \gg 1$), the 321 gauginos become pseudo-Dirac states by 
pairing up with the states that were previously the first KK excited 321 
gauginos.  On the other hand, the XY gaugino states become very light 
in this limit, $m_{\lambda_{\rm XY}} \propto 1/x$ --- it becomes even 
lighter than the MSSM superparticles.  These features are explained in 
more detail in Ref.~\cite{Nomura:2003qb}.

\subsection{4D interpretation}
\label{subsec:4D-interp}

\begin{figure}[t]
\begin{center} 
\begin{picture}(290,95)(5,125)
  \Text(60,145)[t]{\large (a)}
  \Photon(5,190)(32,190){3.5}{3}   \Line(5,190)(32,190)
  \Photon(88,190)(115,190){3.5}{3} \Line(88,190)(115,190)
  \Text(5,198)[b]{$\lambda$} \Text(115,198)[b]{$\lambda$} 
  \GOval(60,190)(23,28)(0){0.85} 
  \Text(61,190)[]{\large SUSY}  \Line(40,185)(80,195)
  \Text(240,145)[t]{\large (b)}
  \DashLine(185,188)(213,188){3} \Text(185,196)[b]{$\tilde{f}$} 
  \DashLine(267,188)(295,188){3} \Text(295,196)[b]{$\tilde{f}$} 
  \PhotonArc(238,188)(25,130,180){3}{2.5} \CArc(238,188)(25,130,180)
  \PhotonArc(242,188)(25,0,50){3}{2.5}    \CArc(242,188)(25,0,50)
  \Text(207,203)[bl]{$\lambda$} \Text(274,203)[br]{$\lambda$}
  \CArc(240,201)(30,204,336) \Text(240,167)[t]{$f$}
  \GOval(240,208)(13,18)(0){0.85} 
  \Text(241,208)[]{\small SUSY}  \Line(225,204)(255,212)
\end{picture}
\caption{Examples of the diagrams that give (a)~gaugino masses and (b)~sfermion 
 masses, where $\lambda$, $\tilde{f}$ and $f$ represent gauginos, sfermions 
 and fermions, respectively.}
\label{fig:gauge-med}
\end{center}
\end{figure}
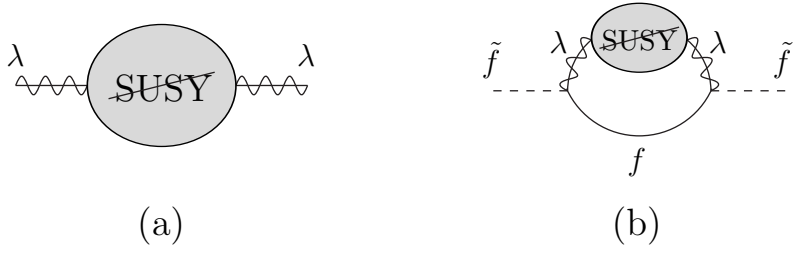

The characteristic features of the spectrum described above can also be 
understood from the 4D picture.  In the 4D picture, the 321 gauginos and 
sfermions obtain masses, for small $x$, from the diagrams as shown in 
Fig.~\ref{fig:gauge-med}.  Here, the gray discs represent contributions 
from the DSB sector.  Using the scaling argument based on the large-$N$ 
expansion, the masses for the gauginos, $M_a \equiv m_{\lambda^{321}_a}$ 
($a=1,2,3$), are estimated as $M_a \simeq g_a^2 (\tilde{N}/16\pi^2) 
\hat{\zeta} m_\rho$, where $g_a$ are the 4D 321 gauge couplings, 
$\hat{\zeta}$ is a dimensionless parameters of $O(1)$, $\tilde{N}$ 
is the size of the DSB gauge group $G$, and $m_\rho$ is the typical 
mass scale for the resonances in the DSB sector.%
\footnote{In a theory where $G$ is almost conformal above the dynamical 
scale $\Lambda$, the parameter $\tilde{N}$ may actually represent the 
{\it square} of the number of ``colors'' of $G$, and not the number of 
``colors'' itself.  Discussions on this and related issues in the AdS/CFT 
correspondence can be found, for example, in Ref.~\cite{Burdman:2003ya}.}
Similarly, the squared masses for the scalars, $m_{\tilde{f}}^2$, are estimated 
as $m_{\tilde{f}}^2 \simeq \sum_{a=1,2,3} (g_a^4 C_a^{\tilde{f}}/16\pi^2) 
(\tilde{N}/16\pi^2) \hat{\zeta}^2 m_\rho^2$, where $\tilde{f} = \tilde{q}, 
\tilde{u}, \tilde{d}, \tilde{l}, \tilde{e}$ represents the MSSM squarks 
and sleptons, and $C_a^{\tilde{f}}$ are group theoretical factors given by 
$(C_1^{\tilde{f}}, C_2^{\tilde{f}}, C_3^{\tilde{f}}) = (1/60,3/4,4/3)$, 
$(4/15,0,4/3)$, $(1/15,0,4/3)$, $(3/20,3/4,0)$ and $(3/5,0,0)$ for $\tilde{f} 
= \tilde{q}, \tilde{u}, \tilde{d}, \tilde{l}$ and $\tilde{e}$, respectively. 

To represent the gaugino and scalar masses in terms of the 5D quantities, 
we use the correspondence relation between the 4D and 5D theories, which are 
given in the present context as $\tilde{N}/16\pi^2 \approx 1/g_B^2 k$ and 
$m_\rho \approx \pi k'$, where $g_B$ represents the $SU(5)$-invariant 5D 
gauge coupling.  The parameter $\hat{\zeta}$ can be read off by matching 
the gaugino mass expressions of 4D and 5D theories as $\hat{\zeta} \approx 
(\zeta g_B^2 F_Z/\pi M_*)$, where the parameter $\zeta$ appears in 
Eq.~(\ref{eq:gaugino-TeV}) and $F_Z$ is the VEV of the highest component 
of the chiral superfield $Z$.%
\footnote{The definition of $F_Z$ here is that, in the normalization where 
the kinetic term of $Z$ is canonically normalized in 4D, $F_Z$ is defined by 
$F_Z = e^{\pi kR} \partial Z/\partial \theta^2|_{\theta = \bar{\theta} = 0}$. 
The natural size for $F_Z$ is then of order $k^2 \sim M_*^2$ (no exponential 
suppression factor).}  Using these relations, we obtain the following 
simple expressions for the gaugino and scalar masses:
\begin{equation}
  M_a = g_a^2\, \frac{\zeta F_Z}{M_*} \frac{k'}{k},
\label{eq:gaugino-masses}
\end{equation}
and 
\begin{equation}
  m_{\tilde{f}}^2 
  = \gamma\!\! \sum_{a=1,2,3} \frac{g_a^4 C_a^{\tilde{f}}}{16 \pi^2}\, 
    (g_B^2 k) \Biggl( \frac{\zeta F_Z}{M_*} \frac{k'}{k} \Biggr)^2,
\label{eq:scalar-masses}
\end{equation}
where $g_a$ are the 4D gauge couplings given by $1/g_a^2 = \pi R/g_B^2 
+ 1/\tilde{g}_{0,a}^2$ ($1/\tilde{g}_{0,a}^2$ are the renormalized 
coefficients for the Planck-brane gauge kinetic terms), and $\gamma$ 
is a numerical coefficient of $O(1)$.  Note that the quantity 
$(\zeta F_Z/M_*)(k'/k)$, appearing in Eqs.~(\ref{eq:gaugino-masses},~%
\ref{eq:scalar-masses}) and setting the overall mass scale, is of $O(M_* 
e^{-\pi kR}/16\pi^2)$, which is naturally of $O(100~{\rm GeV}\!\sim\!1~{\rm 
TeV})$. 

For the case of strong supersymmetry breaking, i.e. $\zeta F_Z/k^2 \gg 1$, 
the 321 gauginos become (pseudo-)Dirac states, where the extra degrees of 
freedom that pair up with the MSSM gauginos arise from the strong $G$ dynamics. 
In this case, the diagram giving the gaugino masses is the one that mixes 
elementary and composite states, instead of Fig.~\ref{fig:gauge-med}a, so 
that the gaugino masses are given by $M_a \simeq g_a (\sqrt{\tilde{N}}/4\pi)\, 
m_\rho \simeq (g_a/\sqrt{g_B^2 k}) (\pi k')$.  The scalar masses are still 
given by the diagram of Fig.~\ref{fig:gauge-med}b, but now with the insertion 
parameter $\hat{\zeta}$ replaced by $1$, as the physics does not depend much 
on the strength of brane supersymmetry breaking in the limit $\zeta F_Z/k^2 
\gg 1$.  This gives $m_{\tilde{f}}^2 \simeq \sum_{a=1,2,3} (g_a^4 
C_a^{\tilde{f}}/16\pi^2) (\tilde{N}/16\pi^2) m_\rho^2 \simeq \sum_{a=1,2,3} 
(g_a^4 C_a^{\tilde{f}}/16\pi^2) (1/g_B^2 k) (\pi k')^2$.  These results, 
together with the formulae in Eqs.~(\ref{eq:gaugino-masses},~%
\ref{eq:scalar-masses}), explain almost all the features observed 
in Refs.~\cite{Goldberger:2002pc,Nomura:2003qb,Chacko:2003tf} for 
the superparticle mass spectrum in warped unified theories.

\subsection{Grand unified particles at colliders}
\label{eq:collider}

In our theory, grand unified particles such as the XY gauge bosons and 
color-triplet Higgs bosons are present at the TeV scale.  What are 
experimental signatures of these particles?  Studying the bulk Lagrangian 
of the theory, we find that it possesses the $Z_2$ parity under which all 
the MSSM states and their KK towers are even while the other ``grand unified 
theory (GUT) states'' are odd: 
\begin{eqnarray}
  && {\rm MSSM\:\: fields}\:\: (+):\:\:\: V_{321}, H_D, Q, L, \cdots, 
\nonumber \\
  && {\rm GUT\:\: fields}\:\: (-):\:\:\:  V_{\rm XY}, H_T, \cdots,
\nonumber
\end{eqnarray}
which we call the GUT parity.  This parity is not broken by the couplings 
present in the theory such as the Yukawa couplings and supersymmetry 
breaking operators.  It can thus be an unbroken symmetry of the theory. 
If this is the case (at least approximately), the lightest GUT particle (LGP) 
is stable at colliders, leading to characteristic experimental signatures.%
\footnote{The GUT parity can in principle be broken by the presence of 
certain brane operators.  In the present case of matter strongly localized 
to the Planck brane, however, the effect of the breaking is suppressed in 
the low-energy 4D theory so that the LGP is still effectively stable for 
collider purposes.  It is possible, however, that the breaking leads to 
the lifetime of the LGP shorter than $\sim 1~{\rm s}$, which may be 
important for cosmology.}

In the theory discussed in section~\ref{sec:warped-sgut}, the LGP is expected 
to be the lightest of the XY gauginos, $\tilde{X}$, and leads to the following 
signatures~\cite{Goldberger:2002pc}.  Because $\tilde{X}$ is colored, 
it will hadronize after production by forming a bound state with a quark 
(or anti-quark).  There are four mesons with almost degenerate masses: 
\begin{equation}
  T^0 \equiv \tilde{X}_{\uparrow}\bar{d}, \quad 
  T^{-} \equiv \tilde{X}_{\uparrow}\bar{u}, \quad 
  T^{\prime -} \equiv \tilde{X}_{\downarrow}\bar{d}, \quad 
  T^{--} \equiv \tilde{X}_{\downarrow}\bar{u}, 
\end{equation}
where $\tilde{X}_{\uparrow}$ and $\tilde{X}_{\downarrow}$ are the isospin 
up and down components of the XY gauginos, respectively.  The mass 
splittings among these states are of order MeV, so that they are all 
sufficiently long-lived to traverse the entire detector without decaying. 
This yields distinctive signals; in particular, the charged states will 
easily be seen by highly ionizing tracks.  These states can also cause 
intermittent highly ionizing tracks, generated through charge/isospin 
exchanges with the detector materials.  The reach of the LHC in the 
masses of these states is estimated to be roughly $2~{\rm TeV}$. 
A detailed analysis for the case of the colored Higgs LGP can be 
found in~\cite{Cheung:2003um}.

\section{Alternative Theories}
\label{sec:alternative}

In this section we present a variety of theories constructed along the 
lines presented in the previous two sections.  The diversity of models 
presented here is an indication of how powerful the framework of warped 
supersymmetric grand unification is, and of the wide variety of phenomena 
we can obtain in this class of theories.

\subsection{321-321 model}
\label{subsec:321-321}

In the model of the previous sections, the bulk $SU(5)$ gauge group is 
broken to 321 on the Planck brane while it is unbroken on the TeV brane. 
We could, however, consider the case where the bulk $SU(5)$ is broken to 
321 both at the Planck and TeV branes.  This class of theories, called 
321-321 theories, was considered in Ref.~\cite{Nomura:2004is}, where it 
was shown that the successful MSSM prediction for gauge coupling unification 
is preserved in such theories.  Specifically, the boundary conditions 
for the bulk fields are given as $V_{321}(+,+)$, $\Sigma_{321}(-,-)$, 
$V_{\rm XY}(-,-)$ and $\Sigma_{\rm XY}(+,+)$ for the gauge sector and 
$H_D,\bar{H}_D(+,+)$, $H_D^c,\bar{H}_D^c(-,-)$, $H_T,\bar{H}_T(-,-)$ and 
$H_T^c,\bar{H}_T^c(+,+)$ for the Higgs sector (realistic theories could 
also be constructed with $H_D,\bar{H}_D(+,-)$, $H_D^c,\bar{H}_D^c(-,+)$, 
$H_T,\bar{H}_T(-,+)$ and $H_T^c,\bar{H}_T^c(+,-)$).  The quark and lepton 
superfields are localized on the Planck brane, and proton decay is adequately 
suppressed.  In the 4D picture, these theories have a similar structure 
to that of the theory in section~\ref{sec:susy-warped} (e.g. that given 
in Fig.~\ref{fig:structure}), but now the global $SU(5)$ symmetry of the 
DSB sector is spontaneously broken at $\Lambda$ by the IR dynamics of $G$. 

As discussed in~\cite{Nomura:2004is}, this class of theories has the 
following distinctive features.
\begin{itemize}
\item Before supersymmetry breaking, the massless sector of the model 
contains not only the MSSM states but also exotic grand unified states 
$\Sigma_{\rm XY}$, $H_T^c$ and $\bar{H}_T^c$ (see Fig.~\ref{fig:KK-321-321}). 
Despite the presence of these exotic states, the successful MSSM prediction 
for gauge coupling unification is preserved.  This is because in the 5D 
picture the wavefunctions of the exotic states are strongly localized to 
the TeV brane so that they do not contribute to the running defined through 
the Planck-brane correlators, and in the 4D picture the exotic states are 
composite and so do not contribute to the running above TeV. 
\begin{figure}[t]
\begin{center} 
\begin{picture}(385,165)(-10,-18)
  \Line(5,0)(355,0) \LongArrow(15,-10)(15,135) 
  \Text(8,135)[r]{mass} \Text(8,75)[r]{$\sim {\rm TeV}$} 
  \Text(45,133)[b]{$V_{321}$}
  \Line(30,0)(60,0)     \Vertex(45,0){3}
  \Line(30,45)(60,45)   \Vertex(45,45){3}
  \Line(30,105)(60,105) \Vertex(45,105){3}
  \Text(85,133)[b]{$\Sigma_{321}$}
  \Line(70,45)(100,45)   \Vertex(85,45){3}
  \Line(70,105)(100,105) \Vertex(85,105){3}
  \Text(125,133)[b]{$V_{\rm XY}$}
  \Line(110,75)(140,75)   \Vertex(125,75){3}
  \Text(165,133)[b]{$\Sigma_{\rm XY}$}
  \Line(150,0)(180,0)   \Vertex(165,0){3}
  \Line(150,75)(180,75)   \Vertex(165,75){3}
  \Text(205,133)[b]{$H_{\!D},\!\bar{H}_{\!D}$}
  \Line(190,0)(220,0)   \Vertex(205,0){3}
  \Line(190,60)(220,60) \Vertex(205,60){3} 
  \Text(245,133)[b]{$H_{\!D}^c,\!\bar{H}_{\!D}^c$}
  \Line(230,60)(260,60) \Vertex(245,60){3} 
  \Text(285,133)[b]{$H_{\!T},\!\bar{H}_{\!T}$}
  \Line(270,90)(300,90) \Vertex(285,90){3} 
  \Text(325,133)[b]{$H_{\!T}^c,\!\bar{H}_{\!T}^c$}
  \Line(310,0)(340,0) \Vertex(325,0){3} 
  \Line(310,90)(340,90) \Vertex(325,90){3} 
\end{picture}
\caption{The schematic picture for the lowest KK spectrum of the 321-321 
 theories before supersymmetry breaking.  After supersymmetry breaking, 
 exotic states of $\Sigma_{\rm XY}$, $H_T^c$ and $\bar{H}_T^c$, as well 
 as the MSSM superparticles, obtain masses of order TeV.}
\label{fig:KK-321-321}
\end{center}
\end{figure}
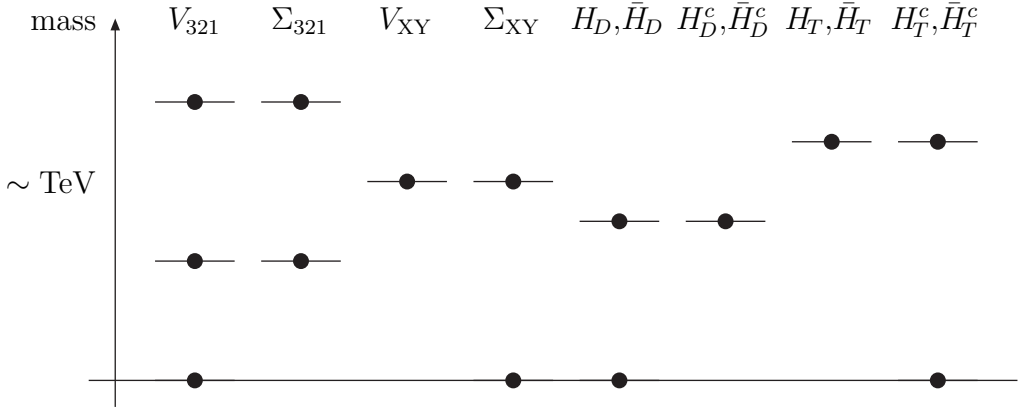
\item After supersymmetry breaking, the exotic states, as well as the MSSM 
superparticles, obtain TeV-scale masses through operators localized on the 
TeV brane.  Because the wavefunctions of the exotic states are strongly 
localized to the TeV brane, while the MSSM states are not, the masses for 
the exotic states are an order of magnitude larger than those of the MSSM 
superparticles.  An exception is the $A_5^{\rm XY}$ state, the fifth 
component of the XY gauge bosons (the imaginary part of the lowest component 
of $\Sigma_{\rm XY}$), whose mass is forbidden at tree level by higher 
dimensional gauge invariance.  The mass of this state is generated at loop 
level so that it could be as light as the MSSM superparticles.  Thus the LGP 
is $A_5^{\rm XY}$, which is expected to be stable at colliders.  In fact, 
we can understand the lightness of $A_5^{\rm XY}$ in the 4D picture, as 
it is the pseudo-Goldstone boson for the breaking $SU(5) \rightarrow 321$ 
caused by the $G$ dynamics.
\item Because supersymmetry is broken at the TeV brane where the gauge 
group is only 321, the generated superparticle masses are non-unified. 
In particular, the masses for the three MSSM gauginos are completely 
free parameters in these theories (because the coupling $\zeta$ in 
Eq.~(\ref{eq:gaugino-TeV}) can take different values for $SU(3)_C$, 
$SU(2)_L$ and $U(1)_Y$).  Squark and slepton masses are also non-unified, 
although they are still flavor universal.  In fact, the superparticle 
masses are given by Eqs.~(\ref{eq:gaugino-masses},~\ref{eq:scalar-masses}) 
with $\zeta$ replaced by $\zeta_a$ (i.e. $\zeta$ now depends on the gauge 
group).   In the 4D picture, we can understand the non-unified feature 
of the spectrum as the result of the spontaneous breakdown of the global 
$SU(5)$ symmetry in the DSB sector. 
\end{itemize}

\subsection{Warped supersymmetric SO(10)}
\label{subsec:SO10}

It is possible to extend the bulk gauge group to a larger unified group 
in warped supersymmetric grand unification.  In particular, we can extend 
the bulk group from $SU(5)$ to $SO(10)$.  The bulk $SO(10)$ is then broken 
to 321 at the Planck brane by a combination of boundary condition and Higgs 
breaking, while at the TeV brane it can either be unbroken or broken to 
the $SU(4)_C \times SU(2)_L \times SU(2)_R$ (422) subgroup.  These theories 
were considered in Ref.~\cite{Nomura:2004it}, and have the following 
(desirable) features.
\begin{itemize}
\item The theories provide an elegant understanding of the matter quantum 
numbers in terms of 422.  In particular, the quantization of hypercharges 
can always be understood no matter how matter fields are introduced. Despite 
the presence of 422, realistic quark and lepton masses and mixings are 
easily obtained through higher dimension operators.
\item Small neutrino masses are obtained quite naturally, as the seesaw 
mechanism arises as an automatic consequence of the theories. 
\item The successful prediction for $b/\tau$ Yukawa unification can 
be reproduced.  The ratio of the VEVs for the two Higgs doublets, 
$\tan\beta \equiv H_D/\bar{H}_D$, is naturally predicted to be large, 
$\tan\beta \approx 50$. 
\item In the case where the gauge group on the TeV brane is 422 and the 
breaking of left-right symmetry is spontaneous, we have a non-trivial 
relation among the three MSSM gaugino masses: $M_1/g_1^2 = (2/5)(M_3/g_3^2) 
+ (3/5)(M_2/g_2^2)$.  This relation still leaves room for the gaugino 
masses to differ from the ones expected from the conventional grand-unified 
mass relations for the gauginos.
\end{itemize}

\subsection{Model with heavy Higgs boson}
\label{subsec:heavy-Higgs}

The construction of warped supersymmetric unification can also be used 
to construct supersymmetric theories in which the mass of the lightest 
Higgs boson is much larger than the conventional upper bound of 
$\approx 130~{\rm GeV}$~\cite{Birkedal:2004zx}.  The basic idea is to 
introduce two sets of Higgs doublets --- one localized on the TeV brane 
receiving a large quartic coupling from the TeV-brane superpotential term 
$W = \lambda S H \bar{H}$, and the other propagating the bulk having the 
Yukawa couplings to the quarks and leptons located on the Planck brane. 
The Higgs doublets responsible for electroweak symmetry breaking are 
linear combinations of these two sets, thus having both the Yukawa couplings 
and a large quartic coupling.  We can then obtain the mass of the lightest 
Higgs boson as large as $\approx 200~{\rm GeV}$, keeping the successful 
MSSM prediction for gauge coupling unification.  This class of theories 
allows the possibility of a significant reduction in the fine-tuning 
needed for correct electroweak symmetry breaking.

\section{Conclusions}
\label{sec:concl}

Supersymmetric unification in warped space provides new possibilities for model 
building.  The picture of warped supersymmetric unification arises naturally 
through the AdS/CFT correspondence from the assumption that supersymmetry 
is dynamically broken at $\Lambda \approx (10\!\sim\!100)~{\rm TeV}$ by 
gauge dynamics $G$ having certain special properties.  In the minimal 
model~\cite{Goldberger:2002pc}, the bulk gauge group is $SU(5)$ broken to 
the 321 subgroup at the Planck brane.  The theory leaves many of the most 
attractive features of conventional unification intact.  In particular, 
the successful MSSM prediction for gauge coupling unification is preserved, 
and small neutrino masses are naturally obtained from the seesaw mechanism. 
Proton decay is also naturally suppressed at a level consistent with 
experiments.  Yet physics at accessible energies could be quite different 
than in the conventional scenario.  The model reveals its higher-dimensional 
nature near the TeV scale, through the appearance of KK towers and an $N=2$ 
supermultiplet structure.  The spectrum of these particles are tightly 
constrained, so that several definite predictions can be drawn with distinct 
experimental signatures.  

I have also presented several variations of the minimal 
model~\cite{Nomura:2004is,Nomura:2004it,Birkedal:2004zx}, which give 
distinct phenomenologies.  These theories differ in the gauge groups of 
the bulk and branes and/or in locations of the Higgs (and matter) fields. 
Taking together, these theories, including the minimal one, provide a 
basis for phenomenological studies of dynamical supersymmetry breaking 
at low energies.

\section*{Acknowledgments}

I would like to thank David Tucker-Smith for reading the manuscript. 
This work was supported in part by the Director, Office of Science, 
Office of High Energy and Nuclear Physics, of the US Department of Energy 
under Contract DE-AC03-76SF00098 and DE-FG03-91ER-40676, by the National 
Science Foundation under grant PHY-0403380, and by a DOE Outstanding 
Junior Investigator award.

\end{document}